\PassOptionsToPackage{dvipsnames}{xcolor}
\documentclass[%
 aip,
 apl,
 amsmath,amssymb,
 reprint,%
]{revtex4-1}

\usepackage{graphicx}% Include figure files
\usepackage{array}[=2016-10-06]
\usepackage{dcolumn}% Align table columns on decimal point
\usepackage{bm}% bold math
\usepackage{orcidlink}

\usepackage[utf8]{inputenc}
\usepackage[T1]{fontenc}
\usepackage{etoolbox}

\usepackage{physics}
\usepackage[normalem]{ulem}

\usepackage{hyperref}% add hypertext capabilities

\definecolor{linkblue}{HTML}{2F9FD8}
\hypersetup{
    colorlinks=true,
    linkcolor=linkblue,
    citecolor=linkblue,
    urlcolor=Blue
}

\makeatletter

\renewcommand{\section}{%
  \@startsection{section}{1}{\z@}%
    {0.6cm}% Space before
    {0.12cm}% Space after
    {\normalfont\small\bfseries\raggedright}%
}

\renewcommand{\subsection}{%
  \@startsection{subsection}{2}{\z@}%
    {0.5cm}% Space before
    {0.12cm}% Space after
    {\normalfont\small\bfseries\raggedright}%
}

\def\cat@comma@active{}

\makeatother

\begin{document}
%%%%%%%%%%%%%%%%%%%%%%%%%%%%%%%%%%%%%%%%%%%%%%%%%%%%%%%%%%%%%
% \preprint{AIP/123-QED}

%%%%%%%%%%%%%%%%%%%%%%%%%%%%%%%%%%%%%%%%%%%%%%%%%%%%%%%%%%%%%
\title[]{
Overcoming configuration bottleneck: Modular pathways to stable semiconductor spin-qubit arrays}
%%%%%%%%%%%%%%%%%%%%%%%%%%%%%%%%%%%%%%%%%%%%%%%%%%%%%%%%%%%%%
\author{Justyna P. Zwolak\orcidlink{0000-0002-2286-3208}}
\email{jpzwolak@nist.gov}
\affiliation{National Institute of Standards and Technology, Gaithersburg, Maryland 20899, USA}%
\affiliation{Joint Center for Quantum Information and Computer Science,
University of Maryland, College Park, Maryland 20742, USA}
\affiliation{Department of Physics, University of Maryland, College Park, Maryland 20742, USA}

\author{Anthony Sigillito\orcidlink{0000-0002-4765-9414}}
\email{asigilli@seas.upenn.edu}
\affiliation{Department of Electrical and Systems Engineering, University of Pennsylvania, Philadelphia, Pennsylvania 19104, USA}%

%%%%%%%%%%%%%%%%%%%%%%%%%%%%%%%%%%%%%%%%%%%%%%%%%%%%%%%%%%%%%
\begin{abstract}
Over the past decade, semiconductor spin qubits have progressed from few-qubit demonstrations toward larger-scale devices fabricated in increasingly reproducible academic and industrial processes. 
This progress marks an inflection point: the central challenge is no longer to demonstrate high-fidelity operation in carefully tuned devices but to discover, verify, and maintain stable operating conditions reliably across many interdependent controls, varied device geometries, and disparate material platforms. 
In this Perspective, we frame spin-qubit operation as a modular automation problem. 
We decompose the workflow into five modules: bootstrapping from minimal prior information, configuration tuning, virtualization of physical gates into effective control axes, qubit-level tuning, and an operation layer with drift-aware maintenance. 
Using recent demonstrations from our work and the broader community, we argue that scalability will depend on explicit interfaces between modules, standardized intermediate data products, and workflow-level metrics such as throughput, success probability, stability time, recovery time, and robustness.
We close by outlining the infrastructure needed to move beyond isolated tuning demonstrations toward sustained operation: qubit-performance-aware feedback, reusable software and benchmark tasks, and tight collaboration among experimental, theoretical, and software efforts.
\end{abstract}
%%%%%%%%%%%%%%%%%%%%%%%%%%%%%%%%%%%%%%%%%%%%%%%%%%%%%%%%%%%%%
\maketitle
%%%%%%%%%%%%%%%%%%%%%%%%%%%%%%%%%%%%%%%%%%%%%%%%%%%%%%%%%%%%%

%%%%%%%%%%%%%%%%%%%%%%%%%%%%%%%%%%%%%%%%%%%%%%%%%%%%%%%%%%%%%
\section{Overview: The inflection point}
\label{sec:intro}
%%%%%%%%%%%%%%%%%%%%%%%%%%%%%%%%%%%%%%%%%%%%%%%%%%%%%%%%%%%%%
Qubits based on spins in quantum dots (QDs) are widely viewed as a promising route to scalable quantum information processing. 
What makes QD platforms distinctive is that their scaling argument is built around manufacturable hardware and reusable control primitives: complementary metal–oxide–semiconductor (CMOS)-aligned fabrication,~\cite{steinacker2025, Laine2025, Ha22-FDQ} coherence in isotopically engineered silicon,~\cite{Yoneda18-SQC} and modular architectures that support routing and shuttling.~\cite{Zwerver2023-lr, Kunne2024-ph, De-Smet2025-ew, van-Riggelen-Doelman2024-uz, White26-EIQ}

In practice, realizing this scaling promise has required a long arc of tuning and control engineering.
While foundational single- and two-qubit control was demonstrated over two decades ago in GaAs,~\cite{Petta05-CMQ} achieving consistently high-fidelity single- and two-qubit operation in silicon required another decade of device engineering, tuning protocols, and control advances.~\cite{Veldhorst2015-mf, Watson18-TQP, Zajac17-RDC, Mills22-HFS, Mills22-TSP, Philips22-UCS, Yoneda18-SQC}
Recent demonstrations now extend these capabilities toward multi-qubit processors and larger integrated arrays.~\cite{Mills22-TSP, Philips22-UCS, Noiri22-FUG, Borsoi22-QCA, John24-TAL, Abraham26-DCQ, Sigillito19-SSC}
This is an \textit{inflection point} for the platform: scaling now exposes tuning, calibration, and stability as system-level constraints, shifting the dominant challenge from isolated high-fidelity operation to the repeatable discovery and maintenance of \textit{stable operating conditions}.~\cite{Zwolak21-AAQ}

%%%%%%%%%%%%%%%%%%%%%%%%%%%%%%%%%%%%%%%%%%%%%%%%%%%%%%%%%%%%%
\begin{figure*}
\includegraphics[width=\textwidth]{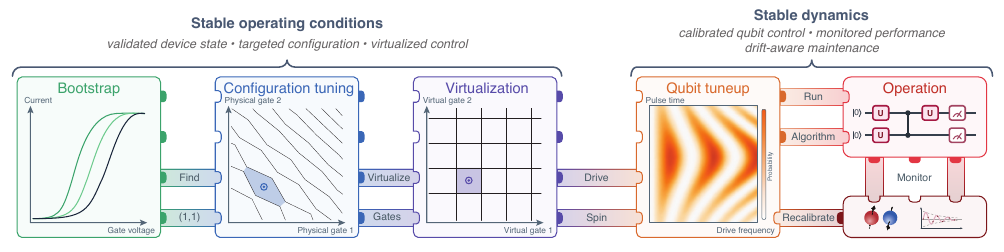}
\caption{\label{fig:modular_stack}
\textbf{From stable operating conditions to stable dynamics.}
Interlocking modules represent an end-to-end workflow for semiconductor spin-qubit operation.
Bootstrapping, configuration tuning, and virtualization establish the device-level operating layer: functional gate ranges and readout conditions, a verified target charge configuration, and effective control axes with reduced cross-capacitance.
These stages provide stable operating conditions for qubit-level tuneup, where spin-control parameters such as drive frequency and pulse duration are calibrated and passed to the operation layer.
During operation, benchmark and drift checks determine whether the system can continue, recalibrate locally, or return to an earlier tuning stage when performance degrades.
}
\end{figure*}
%%%%%%%%%%%%%%%%%%%%%%%%%%%%%%%%%%%%%%%%%%%%%%%%%%%%%%%%%%%%%

In this Perspective, we use stable operating conditions in an explicitly operational sense: a device setting for which the intended charge configuration is realized, and the relevant couplings (to reservoirs, between dots, and to the sensor) lie within functional ranges. 
The operating point must also persist long enough to support repeated experiments without frequent manual intervention, while retaining sufficiently stable dynamics for calibrated control and readout.
Importantly, stability is not a binary attribute but a measurable one, characterized by how long a configuration remains valid, how sensitive it is to small gate offsets or environmental fluctuations, and how rapidly it recovers after perturbation. 
Framed this way, scaling is fundamentally a \textit{configuration problem}: the number of coupled controls grows faster than human intuition, so reproducible operation demands workflows that can find, verify, and maintain stable configurations using standardized, benchmarkable protocols.~\cite{Zwolak21-AAQ, Zwolak23-DNC}

Beyond static operating points, scalable processors require \textit{stable dynamics}: calibrated control and readout sequences must act on qubits in a deterministic, repeatable manner across many experimental cycles.
In practice, this couples configuration stability to qubit calibration because the effective Hamiltonian parameters governing single- and two-qubit operations can be highly sensitive to the local electrostatic environment.
Consequently, even modest drift or miscalibration can translate into appreciable performance loss in multi-qubit operation, motivating workflows that treat tuning and calibration as interdependent steps with explicit verification and recovery.
For spin qubits, which employ interaction-based control (e.g., exchange-mediated operations), small parameter shifts can significantly affect gate performance, so drift-aware recalibration becomes necessary.

As arrays grow, the configuration problem becomes increasingly difficult: the number of coupled control knobs grows faster than the number of reliable, low-cost measurements available to constrain them. 
Even tasks that are ``simple'' in single- and double-dot devices---establishing intended occupancy, locating charge transitions, setting tunnel couplings, and maintaining a sensor working point---become high-dimensional searches because cross-capacitance couples every gate to multiple dot potentials and sensor conditions, while disorder and device-to-device variability reshape the relevant stability regions in ways that are hard to predict \textit{a priori}. 
The operating point is also not static: slow drift, sporadic charge rearrangements, and sensing nonlinearities can move transition features over time, so even a configuration that is ``found'' must be continuously verified and periodically recovered.
Tuning is thus a moving-target control problem: many coupled voltages must be adjusted using sparse, noisy, and time-dependent measurements.
This is precisely the regime where modular automation becomes essential for reproducible scaling.

The case for modularity is reinforced by the diversity of semiconductor spin-qubit platforms and encodings.
The same high-level objective---stable operation---maps to different sensitivity landscapes depending on the material system (e.g., Si/SiGe, Ge/SiGe, GaAs, III-V nanowires, and donor-based devices) and the qubit encoding (e.g., single-spin, singlet-triplet, exchange-only, and hybrid variants).~\cite{Burkard21-SSQ, Chatterjee21-SQP, Stano22-RPM, DeMichielis23-SLI, Severin21-CAT}
These choices affect which control axes matter most, which disorder mechanisms dominate (e.g., charge-offset drift and hysteresis in some gate stacks vs nuclear-spin-induced variability in III–V devices), and which measurements are most informative (or most fragile).
In practice, the dominant constraints can be platform-specific---for example, cross-capacitance and valley physics can narrow viable operating windows in Si/SiGe and Ge/SiGe devices, whereas charge noise, the nuclear-spin environment, and sensor working-point stability can dominate day-to-day operation in GaAs and nanowire platforms.~\cite{Burkard21-SSQ}
Consequently, scalable tuning solutions cannot rely on a single hand-crafted recipe; they require \textit{interoperable modules} that accommodate platform-specific constraints, expose standardized interfaces, and produce intermediate data products suitable for benchmarking, downstream control, and reuse across experimental platforms.

A second practical pressure is that \textit{maintaining} operating conditions is becoming as important as \textit{finding} them.~\cite{Capannelli25-TSD}
Many tuning workflows implicitly assume that a good configuration, once discovered, remains fixed throughout an experiment.
In practice, drift and stochastic reconfiguration degrade operating points on timescales from minutes to days.~\cite{Stewart16-SSE, Hu18-EDD}
What is handled by periodic manual intervention at small scales must, at large scales, become a closed-loop process. 
Routines must automatically detect degradation, localize its cause, and either compensate or reinitialize the device with minimal disruption to uptime.~\cite{Rao25-TFS}
This shifts the tuning paradigm from one-time calibration to \textit{continuous calibration}---an approach familiar from precision instrumentation and increasingly implemented in quantum hardware via \textit{in situ} drift-tracking protocols.~\cite{Evans96-SRR, Wang09-SIL, Kelly16-SIC, Proctor20-DTD}

%%%%%%%%%%%%%%%%%%%%%%%%%%%%%%%%%%%%%%%%%%%%%%%%%%%%%%%%%%%%%
\begin{table*}[t]
\centering
\caption{\label{tab:module_interfaces}
Representative module outputs and metrics for modular semiconductor spin-qubit automation.
Each stage should produce a machine-readable handoff containing the current setting, supporting evidence, validity information, and triggers for recalibration or recovery.
}
\begin{tabular}{p{0.13\textwidth}p{0.28\textwidth}p{0.28\textwidth}p{0.28\textwidth}}
\hline\hline
\textbf{Module} &
\textbf{Data product} &
\textbf{Validation metadata} &
\textbf{Recovery trigger / metric} \\
\hline
Bootstrapping &
Validated gates; operation setpoint; safe bounds &
Gate response; sensor sensitivity; voltage limits; failure flags &
Gate failure; poor sensor response; inaccessible operating window \\
Configuration tuning &
Target charge configuration; transition features; local window &
State confidence; measurement quality; sensor stability &
Wrong regime; degraded sensing; ambiguous transitions; tuning cost \\
Virtualization &
Virtual-gate matrix; compensated axes; residual sensitivities &
Coupling estimates; compensation quality; validity window &
Drifting slopes; loss of compensation; update cost \\
Qubit tuneup &
Resonance frequencies; pulse parameters; readout thresholds &
Calibration timestamp; resonance stability; readout contrast &
Frequency drift; low contrast; calibration time \\
Operation\;and maintenance &
Benchmark records; drift traces; recovery policy &
Benchmark trends; drift indicators; failure annotations &
Stability time; recovery time; uptime \\
\hline\hline
\end{tabular}
\end{table*}
%%%%%%%%%%%%%%%%%%%%%%%%%%%%%%%%%%%%%%%%%%%%%%%%%%%%%%%%%%%%%

These observations motivate the Perspective's organizing question: What does an end-to-end, scalable workflow look like for discovering and maintaining stable operating conditions in semiconductor spin-qubit devices?
We advocate a view in which tuning is decomposed into interoperable modules that can be assembled, replaced, and benchmarked---bootstrapping an uncharacterized device, configuration tuning, virtualization (mapping physical gates onto effective control axes), qubit tuneup and an operation layer with drift-aware maintenance.~\cite{Zwolak21-AAQ}
This modular stack is summarized in Fig.\,\nobreak\ref{fig:modular_stack}, which highlights the transition from stable operating conditions to stable dynamics and the interlocking interfaces required for recovery and reuse.
Much like building with LEGO, scalability depends less on any single ``best'' component than on standardized connectors---interfaces, data products, and metrics---that allow modules to be assembled, replaced, and reused.
Accordingly, progress hinges on clear interfaces between modules, standardized intermediate data products, and workflow-level metrics that quantify performance at scale.~\cite{Zwolak21-AAQ, Zwolak23-DNC, Schuff2026}
In what follows, we develop this view module by module, returning in Sec.\,\nobreak\ref{sec:next_steps} to the infrastructure required to make stable configurations an engineered, maintainable layer of spin-qubit hardware.

%%%%%%%%%%%%%%%%%%%%%%%%%%%%%%%%%%%%%%%%%%%%%%%%%%%%%%%%%%%%%
\section{Towards enabling scalable semiconductor quantum chips}
\label{sec:modular_stack}
%%%%%%%%%%%%%%%%%%%%%%%%%%%%%%%%%%%%%%%%%%%%%%%%%%%%%%%%%%%%%
A scalable approach to operating semiconductor spin qubits requires more than isolated autotuning tricks; it requires an end-to-end workflow with interoperable modules connected by explicit interfaces.
For practical purposes, we decompose this workflow into five stages that can be assembled, tested, and improved independently: (i) \textit{bootstrapping} a new device from minimal prior information to reach a measurement-ready operating point; (ii) \textit{tuning configuration}, which identifies the working regime and reaches the desired charge configuration with quantified confidence; (iii) \textit{virtualization}, which maps physical gate voltages onto effective control axes that enable reproducible adjustments in the presence of cross-capacitance; (iv) \textit{qubit calibration}, which establishes gate-level operation; and (v) \textit{operation with drift-aware maintenance}, which sustains performance via monitoring, targeted recalibration, and recovery.
In Subsections~\ref{ssec:boot}--\ref{ssec:fine}, we summarize each module in terms of its objective, key experimental challenges at scale, and representative strategies, highlighting our contributions where applicable, and using the modular flow depicted in Fig.\,\nobreak\ref{fig:modular_stack} as a guide.

In this stack, the interfaces between modules are as important as the modules themselves.
Each stage must output a machine-readable data product that specifies not only the current device setting, but also the evidence supporting it, the range over which it is expected to remain valid, and the conditions under which the workflow should proceed, repeat a measurement, or return to an earlier stage.
These interface specifications are the ``connectors'' that make modular automation reusable rather than a collection of isolated autotuning routines.
Representative handoffs, metadata, and metrics for each module are summarized in Table\,\nobreak\ref{tab:module_interfaces}.

%%%%%%%%%%%%%%%%%%%%%%%%%%%%%%%%%%%%%%%%%%%%%%%%%%%%%%%%%%%%%
\subsection{Bootstrapping: Initialization from minimal prior knowledge}
\label{ssec:boot}
%%%%%%%%%%%%%%%%%%%%%%%%%%%%%%%%%%%%
Bootstrapping is the first step in the tuning process, aiming to bring an initially uncharacterized device to a measurement-ready operating point from minimal prior information.
This typically includes verifying gate functionality, establishing a sensitive readout condition (e.g., a charge-sensor working point), identifying a viable initial charge regime, and producing a machine-readable starting configuration that downstream modules can consume.
Across platforms, bootstrapping strategies range from staged, physics-informed bring-up recipes with explicit diagnostics~\cite{Darulova19-ATQ, Zubchenko24-ABQ} to adaptive, data-driven workflows that modify measurements and actions in response to observed device behavior.~\cite{Kovach24-BATIS}

As qubit arrays become larger and more strongly coupled,~\cite{Borsoi22-QCA, John24-TAL, George25-SFM, Abraham26-DCQ, Acuna24-CTD} bootstrapping must succeed under complex, nonlinear crosstalk between control lines, microscopic disorder, and time-dependent offsets that reshape the charge landscape during operation.
Consequently, hand-crafted bring-up procedures are difficult to reproduce and even harder to automate reliably across platforms.
From a modular-systems perspective, the key requirement is that bootstrapping output a standardized interface---validated gates, readout setpoints, safe operating bounds, and failure flags---that provides a clean handoff to configuration tuning and virtualization.

Two recent efforts illustrate how this modular viewpoint can be implemented in practice across very different device classes by coupling staged diagnostics to automated analysis and explicit handoff to downstream modules.
\citeauthor{Zubchenko24-ABQ} presents a fully autonomous initialization protocol for a depletion-mode GaAs device that follows a staged, physics-informed bring-up sequence: it validates gate functionality, establishes sensor operation via RF-reflectometry, and hands off a device-ready state to downstream coarse-tuning modules.~\cite{Zubchenko24-ABQ}
The protocol is organized as a sequence of checks and calibrations with explicit success criteria, yielding intermediate outputs (e.g., validated gates and a readout setpoint) that define the handoff to later tuning stages.

BATIS, a bootstrapping, autonomous testing, and initialization system for accumulation-mode Si/SiGe multi-quantum-dot devices, implements a closed-loop adaptive workflow whose executable measurement and action sequence is adjusted in real time based on observed device response.~\cite{Kovach24-BATIS} 
While the high-level \textit{procedural flow} of BATIS is well-defined, the \textit{executable flow} for a particular device---the actual sequence of measurements and operations---is automatically determined by the device architecture and instrument connectivity specified in configuration files.
As tuning progresses, BATIS tailors subsequent diagnostic and calibration stages to match the system's discovered characteristics, including recovery pathways (e.g., illumination) to restore viable operating windows under large random gate-offset shifts induced by trapped charge.
Together, these works illustrate the same modular takeaway across very different device classes: bootstrapping should emit standardized, machine-readable outputs that define a clear starting configuration for the following stages.

%%%%%%%%%%%%%%%%%%%%%%%%%%%%%%%%%%%%%%%%%%%%%%%%%%%%%%%%%%%%%
\subsection{Configuration tuning: Reaching and verifying target configurations}
\label{ssec:state}
%%%%%%%%%%%%%%%%%%%%%%%%%%%%%%%%%%%%
Configuration tuning assumes successful bootstrapping: validated gates, a stable readout setpoint, and an operating window in which charge configurations can be probed reliably. 
The goal is then to infer the current operating regime, identify the relevant charge-transition structure, and drive the device toward a target configuration with sufficient confidence for downstream tuning. 
This target may be expressed coarsely, e.g., a single-dot or double-dot regime, or more specifically as a desired charge occupancy, such as a target $(m,n)$ configuration in a double quantum dot. 

At scale, configuration tuning is difficult because the boundaries between operating regimes are high-dimensional, device-dependent, and subject to drift. 
Cross-capacitance, sensor instabilities, microscopic disorder, and measurement noise can cause the same nominal voltage update to produce different apparent patterns across devices or even across repeated measurements of the same device. 
As a result, small inference errors can compound into large failure rates in closed-loop operation. 
Moreover, reaching the correct charge region is not always sufficient: the optimal working point for qubit operation may not lie at the geometric center of a charge cell but may be skewed by requirements on tunnel coupling, detuning, exchange sensitivity, readout visibility, or microscopic disorder. 
Thus, configuration tuning must both reach a target regime and provide enough local information to support subsequent virtualization and qubit-level refinement.

One strategy for automating configuration tuning is to iterate between ML-enhanced inference and optimizer-driven gate updates until a desired regime is reached.
An early formulation of this approach appears in Ref.~\onlinecite{Kalantre17-MLD}, and it was subsequently demonstrated \textit{in situ} on a Si/SiGe double-QD device.~\cite{Zwolak20-AQD}
In these works, convolutional neural networks (CNNs) trained on simulated data infer the device state from a two-dimensional (2D) scan in plunger-gate space, returning a probability vector over candidate states that can be used as feedback for navigation in gate space.
To reduce acquisition overhead, a ray-based classification (RBC) representation was introduced in which state information is extracted from a structured set of evenly distributed one-dimensional (1D) traces (rays) centered about a point.~\cite{Zwolak21-RBI}
By replacing dense raster scans with targeted 1D probes and a physics-informed navigation strategy,~\cite{Ziegler22-TAR} RBC enables state identification and closed-loop tuning with substantially fewer measurements than conventional 2D scans.

A complementary view treats configuration tuning as a charge-setting problem: starting from a known regime, the device is emptied and then loaded to the target occupancy by detecting and counting charge transitions.
At a high level, such workflows combine \textit{structured measurements} that expose transitions, \textit{inference} that extracts actionable features such as transition locations and slopes, and \textit{control updates} that move the device toward a specified occupancy.
The first automated protocol for charge tuning combined ML inference from 2D measurements with an unloading-and-reloading strategy.~\cite{Durrer19-ATQ}
A subsequent measurement-efficient approach leveraged transition geometry from 1D probes to guide gate updates and used local cross-capacitance compensation during the reloading stage.~\cite{Ziegler22-TAR}

This line of work underscores both the promise and the practical challenges of ML-driven automation.
Reliable configuration tuning requires not only accurate state labels or transition locations but also confidence that the underlying measurement is sufficiently informative to support the next closed-loop action.
In practice, this means accounting for model limitations, sensor stability across repeated measurements, and the sensitivity of optimization to initialization and to nonconvex or ill-conditioned objectives.
A robust workflow, therefore, needs an explicit safeguard against acting on degraded or misleading data.
One example is a ``gatekeeper'' module that detects degraded sensing conditions and triggers re-measurement, recalibration, or reinitialization rather than silently propagating errors.~\cite{Ziegler22-TRA}
The modular output of configuration tuning is therefore a verified target configuration that carries confidence information, local transition features, and an operating window (Table\,\nobreak\ref{tab:module_interfaces}), which is passed to virtualization and qubit calibration.

%%%%%%%%%%%%%%%%%%%%%%%%%%%%%%%%%%%%%%%%%%%%%%%%%%%%%%%%%%%%%
\subsection{Virtualization: From physical gates to effective control axes} 
\label{ssec:virt}
%%%%%%%%%%%%%%%%%%%%%%%%%%%%%%%%%%%%
Configuration tuning is most effective when the control space is well parameterized.
In gate-defined QD devices, the physical control voltages are generally not aligned with the parameters one would like to tune: changing a single gate can shift multiple dot chemical potentials, alter sensor response, and perturb interdot tunnel couplings.
The goal of virtualization is to transform this dense, cross-coupled physical gate space into an effective control basis in which key Hamiltonian parameters can be adjusted approximately, independently, and reproducibly.
At the electrostatic level, this often means defining \textit{virtual gates} as prescribed linear combinations of physical gates so that one can move individual dot chemical potentials, compensate sensor response, or adjust barriers while minimizing unintended changes elsewhere in the device.

As devices grow, the number of relevant cross-capacitances and operating regimes increases, requiring efficient calibration, continuous validation, and updates as the operating point shifts. 
Virtualization is commonly achieved by extracting relative capacitive couplings from charge-sensing measurements and using them to construct virtual control axes.~\cite{Che2024, Baart16-CAT, Mills19-CAT}
In practice, scalable solutions benefit from layering this process so that earlier transformations, such as sensor compensation, are automatically preserved as higher-level axes are constructed.
An illustrative example is the modular autonomous virtualization system (MAViS), which operationalizes virtualization as a \textit{stack} of interoperable calibration layers that can be executed and validated autonomously in real time.~\cite{Rao24-MAViS}
MAViS combines ML-based feature extraction from small charge-stability diagrams with computer-vision and regression tools to determine the relative couplings needed to orthogonalize and normalize plunger control and to construct virtual barrier controls that tune interdot coupling while maintaining the charge configuration.

This electrostatic view of virtualization provides a critical interface between configuration tuning and qubit-level calibration, but it is not the end of the problem.
Much of the early work on virtualization has focused on electrostatic parameters---dot chemical potentials, sensor compensation, and tunnel-barrier controls---because these govern the charge dynamics required to form and stabilize quantum-dot arrays.
Virtualizing these parameters enables reproducible motion through the charge space and provides the control basis needed to access regimes in which the array can host qubits.~\cite{Hensgens17-FHQ, Hsiao20-EOT, Qiao20-CME}
However, the next, more challenging step is to extend the same philosophy to \emph{qubit dynamics}: effective controls must account not only for charge-state stability but also for crosstalk that affects single-qubit rotations, exchange interactions, and multi-qubit entangling operations.

The difficulty is that spin-relevant controls often depend on microscopic and device-specific physics that are only indirectly visible in charge-sensing measurements.
For example, electric-dipole spin resonance (EDSR) in hole-spin qubits can rely on spin--orbit-mediated $g$-tensor modulation, whereas EDSR in electron-spin qubits coupled to micromagnets converts electrically driven motion into an effective ac magnetic field through a magnetic-field gradient.
Consequently, effective lever arms for spin control can depend on spin--orbit coupling, magnetic-field gradients, nuclear or magnetic disorder, and temperature-dependent changes in the local environment.~\cite{Crippa18-ESD, Undseth23-UTD, Burkard21-SSQ}
Similarly, tunnel coupling is related to exchange, but charge-sensing features used to estimate tunnel coupling provide only an indirect and thermally broadened proxy for the Hamiltonian parameters relevant to high-fidelity multi-qubit gates.~\cite{DiCarlo04-DCS}
Qubit-based measurements, including exchange spectroscopy and dynamically decoupled exchange oscillations, can then be used to validate and refine the control model at the level of the relevant qubit dynamics.~\cite{Madzik21-CQO, Connors22-CNS}
Thus, stabilizing the charge configuration can inadvertently perturb spin control, and charge-state recalibration will often need to be followed by a refinement layer based directly on qubit-performance measurements.

The modular output of virtualization is therefore more than a virtual-gate matrix. 
It should carry a validity window, a record of which physical parameters are compensated, and the residual sensitivities handed off to qubit-level fine-tuning (Table\,\nobreak\ref{tab:module_interfaces}). 
In this sense, virtualization is both a device-tuning module and a bridge: it translates configuration tuning into a control basis suitable for calibration while making explicit where spin-based measurements and drift-aware maintenance must take over.

%%%%%%%%%%%%%%%%%%%%%%%%%%%%%%%%%%%%%%%%%%%%%%%%%%%%%%%%%%%%%
\begin{figure*}
\includegraphics[width=\textwidth]{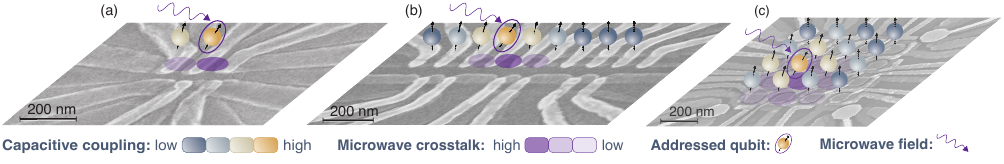}
\caption{\label{fig:device_complexity}
\textbf{Architecture-dependent local control complexity in semiconductor spin-qubit arrays.}
As quantum-dot architectures grow in size and complexity, differences in device geometry,  connectivity, and control crosstalk introduce additional constraints on local tuning and calibration.
The blue-orange shading schematically represents electrostatic cross-capacitance, whereby a voltage change on one gate shifts the chemical potentials of multiple dots.
The purple shading illustrates the microwave-drive footprint, representing the spatial extent of the oscillating control field and its potential to drive neighboring qubits through electrical crosstalk or nonlocal field coupling.
The shading is schematic and does not imply a systematic increase in the range or strength of crosstalk with array size.
Unlike DC cross-capacitance, which shifts the electrostatic operating conditions, microwave-drive crosstalk can induce unintended spin rotations or other dynamical responses, depending on the frequency and amplitude of the drive and the susceptibility of neighboring qubits.
Addressed qubits are indicated with purple ovals.
The overlap between control fields, neighboring dots, and sensing regions illustrates how device geometry and connectivity complicate local tuning and highlights the need for virtualized controls, explicit validity windows, crosstalk-aware spin calibration, and drift-aware monitoring.
(a) A double-QD device adapted from Ref.~\onlinecite{Zajac17-RDC}. 
(b) A nine-QD array adapted from Ref.~\onlinecite{Zajac16-SGA}.
(c) A two-dimensional 16-QD array adapted from Ref.~\onlinecite{Borsoi22-QCA}.
}
\end{figure*}
%%%%%%%%%%%%%%%%%%%%%%%%%%%%%%%%%%%%%%%%%%%%%%%%%%%%%%%%%%%%%

%%%%%%%%%%%%%%%%%%%%%%%%%%%%%%%%%%%%%%%%%%%%%%%%%%%%%%%%%%%%%
\subsection{Fine-tuning: Qubit calibration and drift-aware maintenance}
\label{ssec:fine}
%%%%%%%%%%%%%%%%%%%%%%%%%%%%%%%%%%%%
This subsection combines the final two modules of the workflow: qubit-level tuneup, which establishes calibrated control and readout parameters, and operation, which monitors those parameters and triggers recalibration or recovery when performance degrades.
Fine-tuning begins once a device has been brought to a verified charge configuration and an effective control basis has been established.
At this stage, the goal is to calibrate qubit-relevant parameters---readout and control setpoints, exchange or detuning operating points, resonance frequencies, and gate pulses---and then maintain performance autonomously in the presence of drift.
In contrast to earlier stages, success is no longer defined only by charge stability or transition geometry, but by measured qubit performance: readout contrast, coherence, gate fidelity, exchange stability, and the reproducibility of calibrated dynamics.

It is important to note that drift does not occur only at the device level; classical control electronics are also subject to noise and drift. 
Reliable qubit calibration therefore requires calibration of the classical control and measurement infrastructure, including cable transmission characteristics, digital-to-analog converter offsets, magnetic-field settings, and microwave modulators. 
Imperfections or drift in these components can introduce control errors that must be characterized and compensated to ensure reproducible qubit operation.

A central challenge is that spin-qubit parameters can be highly sensitive to small voltage changes and microscopic details of the device environment.
For example, the Stark shift---the dependence of qubit frequency on applied voltage---can vary strongly from dot to dot and may differ in magnitude, sign, or functional form depending on the local confinement, spin--orbit coupling, micromagnet field gradients, valley physics, and disorder.~\cite{Hanson07-SQD}
Even in micromagnet-free silicon quantum dots, Stark shifts and $g$-factors can vary from dot to dot because of microscopic interface disorder, including monatomic steps at the Si/SiGe or Si/SiO$_2$ interface.~\cite{Ferdous18-ISO}
As a result, stabilizing the charge configuration is necessary but not sufficient: voltage corrections that preserve charge occupation may still shift resonance frequencies, alter exchange and readout visibility, or introduce crosstalk between neighboring qubits.
Figure\,\nobreak\ref{fig:device_complexity} illustrates how device geometry and connectivity shape the local control landscape, with electrostatic and microwave coupling introducing architecture-dependent crosstalk that must be accounted for during qubit calibration.
Fine-tuning, therefore, requires feedback not only from charge-sensing diagnostics but also from qubit-level measurements.

The calibration problem also becomes increasingly coupled as systems scale.
Measuring spin parameters generally requires repeated qubit-level experiments and background calibration, which introduces overhead compared with fast charge-sensing diagnostics.
Moreover, voltage updates applied to one qubit can perturb the operating conditions of neighboring qubits through electrostatic crosstalk, shared sensing circuitry, or control-line coupling.~\cite{Capannelli25-TSD, Cifuentes24-IEC}
Different qubits in the same processor can respond differently to the same class of voltage correction, because Stark shifts and $g$-factor sensitivities depend on local interface disorder, magnetic-field orientation, and device-specific electrostatics.~\cite{Ferdous18-ISO, Cifuentes24-IEC}
As arrays move beyond small demonstrations, scalable calibration will require strategies that identify the most informative qubit-level measurements, update coupled control parameters with minimal overhead, and decide when local recalibration is sufficient vs when earlier modules must be revisited.

Fine-tuning is therefore not a terminal step but the entry point to drift-aware maintenance.
Lightweight ``health checks'' can monitor signatures of degradation---such as readout contrast, resonance shifts, exchange drift, leakage, or changes in sensor response---and trigger targeted recalibration before performance loss propagates through the workflow.
A recent step in this direction is the time-calibration framework demonstrated in Ref.~\onlinecite{Rao25-TFS}, which uses repeated charge-stability measurements as a multidimensional probe of the electrostatic environment, tracking charge-transition motion in time to detect drift, identify abrupt charge reconfigurations, apply compensating voltage updates, and characterize dot-specific noise processes in a 10-QD device.
The modular output of this final stage is not merely a calibrated pulse set but a maintenance policy: calibrated operating parameters, validity windows, recheck schedules, and recovery actions that determine how the system remains usable over time.
This closes the loop in the modular tuning stack, turning stable configurations from static endpoints into operating conditions that can be autonomously monitored, repaired, and maintained.

%%%%%%%%%%%%%%%%%%%%%%%%%%%%%%%%%%%%%%%%%%%%%%%%%%%%%%%%%%%%%
\section{Beyond tuning: Engineering sustained spin-qubit operation}
\label{sec:next_steps}
%%%%%%%%%%%%%%%%%%%%%%%%%%%%%%%%%%%%%%%%%%%%%%%%%%%%%%%%%%%%%
The modular stack reframes tuning as an engineering problem: not the one-time discovery of a favorable operating point, but the repeated discovery, verification, and maintenance of \textit{stable operating conditions}. 
The objective is not to find \textit{any} working configuration, but one that remains compatible with high-fidelity control and readout, is robust to modest perturbations in gate offsets and device environment, and is recoverable when drift or stochastic reconfiguration occurs.
As arrays grow, the number of viable configurations may not increase; usable regions often become more constrained, more fragile, or more dependent on correlations among many controls.
This is where the next generation of automation tools must operate.

%%%%%%%%%%%%%%%%%%%%%%%%%%%%%%%%%%%%%%%%%%%%%%%%%%%%%%%%%%%%%
\subsection{From device tuning to quantum-control maintenance}
%%%%%%%%%%%%%%%%%%%%%%%%%%%%%%%%%%%%
A first priority is to move beyond charge-configuration automation toward qubit-level calibration and maintenance. As established above, a workflow that stabilizes charge occupation but does not monitor qubit dynamics is incomplete: control parameters, amplitudes, frequencies, durations, phases, operating points, and readout thresholds are qubit- and gate-dependent, drift over time, and couple across the device through electrostatic, control-line, and frequency crosstalk.~\cite{Burkard21-SSQ, Vahapoglu22-CCQ, Unseld25-BCQ, Capannelli25-TSD, Cifuentes24-IEC, Undseth23-UTD}
Echoing a broader trend across quantum hardware,~\cite{Kelly16-SIC, Proctor20-DTD} future automation must close the loop on measured qubit performance, using spin-based spectroscopy, Ramsey-style frequency tracking, exchange calibration, randomized benchmarking, or related diagnostics to decide when a local correction is sufficient and whether the device must be returned to an earlier stage of the tuning stack.

%%%%%%%%%%%%%%%%%%%%%%%%%%%%%%%%%%%%%%%%%%%%%%%%%%%%%%%%%%%%%
\subsection{From isolated demonstrations to engineered workflows}
%%%%%%%%%%%%%%%%%%%%%%%%%%%%%%%%%%%%
A second priority is to recognize that the path to scale will differ across research environments and technology strategies.
Historically, progress in semiconductor spin qubits has been limited not only by control concepts but also by the difficulty of reliably producing working multi-dot devices.
The development of the overlapping-gate architecture enabled reproducible few-qubit devices and multi-qubit demonstrations in academic laboratories.~\cite{Zajac16-SGA, Philips22-UCS, Noiri22-FUG, Borsoi22-QCA}
In parallel, increasingly industrialized fabrication flows, including etched architectures,~\cite{Ha22-FDQ} wafer-scale electron-beam lithography,~\cite{Elsayed24-LCQ} and advanced CMOS processing,~\cite{Neyens24-PQW} are enabling larger and more reproducible device arrays.~\cite{Nickl25-EQO, George25-SFM, Abraham26-DCQ}
Academic laboratories remain especially well-positioned to explore new materials, gate architectures, qubit encodings, sensing modalities, and control algorithms, including disruptive ideas that may not yet be manufacturing-ready.
Industrial and national laboratory efforts can complement this role by driving improvements in yield, reproducibility, cryogenic integration, wiring density, process control, and system-level engineering across many nominally similar devices. 
Both roles are essential.

This division of effort is not only productive, but also changes the tuning problem.
As hardware matures, the field is moving beyond simple Loss--DiVincenzo-style single-spin operation toward a broader design space that includes shuttling architectures,~\cite{Zwerver2023-lr, Kunne2024-ph, De-Smet2025-ew} sparse-readout layouts,~\cite{Boter22-SWA, Kammerloher24-SDS} and multi-dot encodings designed to reduce sensitivity to selected noise sources or simplify particular control requirements.~\cite{Ha22-FDQ, Weinstein22-ULS}
These design choices can reduce one bottleneck while introducing another: encoded qubits may suppress sensitivity to charge or magnetic-field noise, but they often require more elaborate charge configurations, additional tunnel-coupling calibration, and tighter coordination among multiple gates and sensors.
Similarly, architectures that reduce reliance on microwave control, such as exchange-only approaches, can shift complexity from microwave delivery to exchange calibration and electrostatic stability.~\cite{DiVincenzo00-UQE, Weinstein22-ULS}
Conversely, highly engineered fabrication can reduce variability, but it does not eliminate the need for autonomous workflows; larger, denser, and more interconnected arrays make manual configuration and maintenance increasingly impractical.

Recent industrial demonstrations illustrate the importance of integrated automation for scaling semiconductor spin-qubit processors. 
For example, the silicon quantum processing unit demonstrated in Ref.~\onlinecite{HRLQT26-DCQ} combines device tuning, qubit calibration, and cryogenic control electronics within an integrated architecture, and comparable integrated calibration infrastructures underpin the operation of large superconducting processors.\cite{Kelly16-SIC, Arute19-QSS} 
Such advances demonstrate how integrated automation can support the operation of increasingly complex quantum processors.
However, achieving reliable automated operation within a particular hardware architecture is distinct from developing automation methods that can be readily transferred between devices, experimental platforms, and research groups. 
Addressing this second challenge requires interoperable software components, standardized interfaces, and reusable representations of tuning and calibration procedures.

%%%%%%%%%%%%%%%%%%%%%%%%%%%%%%%%%%%%%%%%%%%%%%%%%%%%%%%%%%%%%
\subsection{From algorithms to infrastructure}
%%%%%%%%%%%%%%%%%%%%%%%%%%%%%%%%%%%%
A third priority is to build the software infrastructure needed to meaningfully compare automation methods.~\cite{Kovach26-FALcon}
Algorithmic tuning demonstrations have increased in number, but they remain difficult to compare because they often target different devices, assume different prior information, use different measurements, and report different success criteria.~\cite{Zwolak21-AAQ}
The community would benefit from shared benchmark tasks that reflect realistic workflows: bootstrapping from minimal prior information, reaching a target configuration under measurement constraints, maintaining a configuration under drift, and recovering after abrupt charge rearrangements.
Equally important are standardized data products---charge maps with metadata, transition labels and confidence scores, virtual-gate matrices with validity windows, drift traces, calibration logs, and failure annotations---that allow methods to be tested and improved outside the laboratories where they were developed.
Such benchmarks would not replace device-specific expertise, but they would quantify progress.

Beyond benchmarking, interoperability requires separating the logic of tuning and calibration algorithms from their device- and instrument-specific implementations. 
This separation allows high-level workflows to be adapted to different hardware architectures without requiring identical physical control mechanisms or measurement procedures. 
Developing software infrastructure around standardized interfaces and reusable workflow descriptions can facilitate the exchange and transfer of automation methods across experimental platforms and research groups.~\cite{Kovach26-FALcon}

Meeting these goals will require tight collaboration among experimentalists, theorists, software developers, and standards-oriented measurement scientists.
Experiments define the relevant failure modes and provide the feedback signals; theory and modeling identify the physical structure that makes learning data-efficient; software engineering turns algorithms into robust, reusable tools; and benchmarking connects local progress to community-wide assessment.
This collaboration is especially important because scalable automation will not be achieved by a single universal algorithm.
Rather, it will require interoperable modules with clear assumptions, explicit handoffs, and metrics that quantify not only whether tuning succeeds but also how long it takes, how often it fails, how much data it consumes, how stable the resulting operation is, and how quickly the system recovers.

The central challenge for the field is therefore to turn stable operating conditions into an engineered layer of semiconductor spin-qubit hardware.
This requires not only better tuning modules, but also well-defined connectors---interfaces, data products, and metrics---that enable those modules to be assembled into reliable operating workflows.
In this view, the central object of automation is not a voltage vector or a trained model, but a verified operating state together with the evidence, validity window, and recovery policy needed to keep it useful.
If semiconductor spin qubits are to move from exquisitely tuned laboratory demonstrations to scalable, manufactured quantum systems, this infrastructure must become reliable, measurable, and portable across devices, laboratories, and platforms.

%%%%%%%%%%%%%%%%%%%%%%%%%%%%%%%%%%%%%%%%%%%%%%%%%%%%%%%%%%%%%
\begin{acknowledgments}
This research was sponsored in part by the U.S. Army Research Office (ARO) under Award Nos. W911NF-23-1-0258 and W911NF-24-2-0043.
The views and conclusions contained in this paper are those of the authors and should not be interpreted as representing the official policies, either expressed or implied, of the U.S. Government. 
The U.S. Government is authorized to reproduce and distribute reprints for Government purposes, notwithstanding any copyright noted herein. 
Any mention of commercial products is for informational purposes only; it does not imply recommendation or endorsement by the National Institute of Standards and Technology.
\end{acknowledgments}
%%%%%%%%%%%%%%%%%%%%%%%%%%%%%%%%%%%%%%%%%%%%%%%%%%%%%%%%%%%%%

%%%%%%%%%%%%%%%%%%%%%%%%%%%%%%%%%%%%%%%%%%%%%%%%%%%%%%%%%%%%%
\section*{Author declarations}
%%%%%%%%%%%%%%%%%%%%%%%%%%%%%%%%%%%%%%%%%%%%%%%%%%%%%%%%%%%%%
\subsection*{Conflict of Interest}
%%%%%%%%%%%%%%%%%%%%%%%%%%%%%%%%%%%%
The authors have no conflicts to disclose.

%%%%%%%%%%%%%%%%%%%%%%%%%%%%%%%%%%%%%%%%%%%%%%%%%%%%%%%%%%%%%
\subsection*{Author Contributions}
%%%%%%%%%%%%%%%%%%%%%%%%%%%%%%%%%%%%
\noindent\textbf{Justyna P. Zwolak}: Conceptualization (lead); Funding acquisition (lead); Project administration (equal); Visualization (equal); Writing -- original draft (lead); Writing -- review \& editing (equal). 
\textbf{Anthony Sigillito}: Conceptualization (supporting); Project administration (supporting); Visualization (equal); Writing -- original draft (supporting); Writing -- review \& editing (equal).

%%%%%%%%%%%%%%%%%%%%%%%%%%%%%%%%%%%%%%%%%%%%%%%%%%%%%%%%%%%%%
\section*{Data Availability Statement}
%%%%%%%%%%%%%%%%%%%%%%%%%%%%%%%%%%%%%%%%%%%%%%%%%%%%%%%%%%%%%
Data sharing is not applicable to this article as no new data were created or analyzed in this study.

%%%%%%%%%%%%%%%%%%%%%%%%%%%%%%%%%%%%%%%%%%%%%%%%%%%%%%%%%%%%%
%merlin.mbs aipnum4-1.bst 2010-07-25 4.21a (PWD, AO, DPC) hacked
%Control: key (0)
%Control: author (8) initials jnrlst
%Control: editor formatted (1) identically to author
%Control: production of article title (-1) disabled
%Control: page (0) single
%Control: year (1) truncated
%Control: production of eprint (0) enabled
%
%%%%%%%%%%%%%%%%%%%%%%%%%%%%%%%%%%%%%%%%%%%%%%%%%%%%%%%%%%%%%

%%%%%%%%%%%%%%%%%%%%%%%%%%%%%%%%%%%%%%%%%%%%%%%%%%%%%%%%%%%%%
\end{document}